%% file: main.tex
\documentclass[lettersize,journal]{IEEEtran}
\usepackage{amsmath,amsfonts,amssymb}
\usepackage{algorithmic}
\usepackage{algorithm}
\usepackage{array}
\usepackage{subcaption}
\usepackage{textcomp}
\usepackage{flushend}
\usepackage{url}
\usepackage{verbatim}
\usepackage{bbm}
\usepackage{dsfont}
\usepackage{nicefrac,xfrac}
\usepackage{graphicx}
\usepackage[noadjust]{cite}
\usepackage{color}

\begin{document}

\title{Improving the Spectral Efficiency of \\Zak-OTFS via Mutually Unbiased Bases}

\author{Sandesh Rao Mattu$^*$, Nishant Mehrotra$^*$, Robert Calderbank~\IEEEmembership{Fellow,~IEEE}\vspace{-7mm}
\thanks{This work is supported by the National Science Foundation under grants 2342690 and 2148212, in part by funds from federal agency and industry partners as specified in the Resilient \& Intelligent NextG Systems (RINGS) program, and in part by the Air Force Office of Scientific Research under grants FA 8750-20-2-0504 and FA 9550-23-1-0249. \\ The authors are with the Department of Electrical and Computer Engineering, Duke University, Durham, NC, 27708, USA (email: \{sandesh.mattu,\allowbreak nishant.mehrotra,\allowbreak robert.calderbank\}\allowbreak@duke.edu). \\ $*$ denotes equal contribution.}
}



\maketitle

\begin{abstract}
Orthogonal signaling or Nyquist signaling limits the number of information symbols transmitted in bandwidth $B$ and time $T$ to be $BT$, the time-bandwidth product. Transmitting more than $BT$ symbols leads to loss of orthogonality. 
The standard approach is to reduce the symbol interval and resolve the resulting inter-symbol interference.
This requires changing the
sampling frequency and possibly the sampling clock. 
This paper shows that it is possible to improve spectral efficiency on doubly spread channels without changing the sampling frequency. 
The idea is to superimpose the information symbols using mutually unbiased bases (MUB) while maintaining the original spacing. We carry this out in the delay-Doppler domain using Zak-transform based orthogonal time frequency space (Zak-OTFS) modulation as it allows construction of MUB. We also construct a precoder that mitigates the effect of the doubly-spread channel. This simplifies receiver processing to detection in Gaussian noise since each basis appears to the other as Gaussian noise. This reduction makes it possible to use trellis coded modulation to further improve the bit-error performance. Numerical results demonstrate that the proposed signaling scheme using MUB achieves good bit-error performance.
\end{abstract}

\begin{IEEEkeywords}
DD domain, mutually unbiased bases, precoder, TCM, Zak-OTFS. 
\vspace{-2mm}
\end{IEEEkeywords}

\input{intro}

\input{sys_model}

\input{prop_method}

\input{results}

\input{conclusion}

\bibliographystyle{IEEEtran}
\bibliography{references}

\end{document}

%% file: intro.tex
\section{Introduction}
\label{sec:intro}
\IEEEPARstart{D}{ata} transmission in bandwidth-limited channels, which are characteristic of wireless and wireline communication systems, is typically performed at the Nyquist rate. Formally, for a channel with bandwidth $B$, the number of independent data symbols that can be transmitted at the Nyquist rate in a time interval $T$ is $BT$~\cite{Tse2005}. This is achieved by mounting $BT$ independent information symbols on time-frequency pulses with the property that shifts in time of $\nicefrac{1}{B}$ (or its multiples) and shifts in frequency of $\nicefrac{1}{T}$ (or its multiples) are orthogonal to one another. In other words, the information symbols are mounted on a $BT$-dimensional basis which spans the $BT$ space. This concept of Nyquist signaling is also easily extended to communication in signal domains other than time-frequency, e.g., to the delay-Doppler (DD) domain, as exemplified by Zak transform based orthogonal time frequency space, or Zak-OTFS~\cite{bitspaper1}\cite{bitspaper2}, by defining pulses that are orthogonal to shifts in delay of $\nicefrac{1}{B}$ and shifts in Doppler of $\nicefrac{1}{T}$ (or their multiples). The idea remains to mount $BT$ information symbols on an orthogonal basis defined in the appropriate domain.

\begin{figure}
    \centering
    \begin{subfigure}{\linewidth}
        \includegraphics[width=\linewidth]{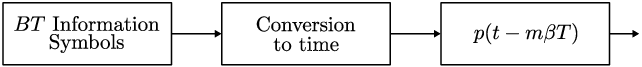}
        \caption{Conventional (Faster-than-Nyquist signaling)}
        \label{fig:conv_approach}
    \end{subfigure}\\
    \vspace{2mm}
    \begin{subfigure}{\linewidth}
        \includegraphics[width=\linewidth]{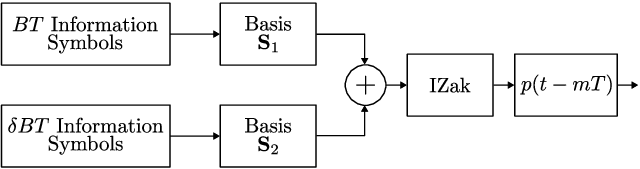}
        \caption{MUB}
        \label{fig:mub_approach}
    \end{subfigure}
    \caption{Comparison between two ways of improving spectral efficiency. Conventional scheme improves spectral efficiency by increasing the sampling frequency (captured by ${0 < \beta < 1}$), while the MUB approach uses mutually unbiased bases to mount $BT + \delta BT = BT(1+\delta), 0<\delta<1$ information symbols, which requires no change in the sampling rate ($\beta = 1$). The spectral efficiency of both the approaches is related as $\nicefrac{1}{\beta} = (1+\delta)$.}
    \label{fig:ftn_comp}
    \vspace{-5mm}
\end{figure}

Although the maximum number of orthogonal information signals that are possible in a $BT$-dimensional space is $BT$, more than $BT$ information symbols can be transmitted if the interference from non-orthogonal signaling can be handled at the transmitter/receiver. This forms the basis for faster-than-Nyquist signaling~\cite{mazo_ftn, liveris_ftn}. Most of the schemes achieving higher spectral efficiency~\cite{paul2025cholesky,rusek_ftn, calavolpe_ftn_tf_2lattice_poweralloc_turbo, zhang_ftn_tf_hexalattice_isioptmzn} use the setup shown in Fig.~\ref{fig:ftn_comp}(\subref{fig:conv_approach}). In this approach, the information symbols in the transform domain are converted to time domain and mounted on a band-limited pulse which is sampled at $\beta T, 0<\beta<1$, where $T$ is the Nyquist sampling interval. This requires changing the sampling frequency and possibly the sampling clock. 
Changing the sampling frequency introduces inter-symbol interference (ISI).

In this paper we take the mutually unbiased bases (MUB) approach shown in Fig~\ref{fig:ftn_comp}(\subref{fig:conv_approach}). Two bases are said to be MUB if the interference from one basis looks like Gaussian noise to the other. We construct two bases $\mathbf{S}_1$ and $\mathbf{S}_2$ so that they are mutually unbiased (see~\eqref{eq:mut_unb_basis}). This construction is carried out in the delay-Doppler (DD) domain~\cite{Aug2024paper, preamblepaper} using Zak-transform based orthogonal time frequency space (Zak-OTFS) modulation~\cite{bitspaper1, bitspaper2}. $BT$ information symbols are mounted on the first basis and $\delta BT, 0 < \delta < 1$ symbols are mounted on the second basis. The modulated bases are then superimposed. The inverse Zak transform block converts the DD symbols to time which is mounted on a pulse sampled at the \textit{Nyquist rate}. Effectively, in MUB, the interference is moved from faster sampling of the transmit pulse to the interference between the two bases.

The proposed approach of using MUB with Zak-OTFS for improving the spectral efficiency has another advantage. The channel in the DD domain remains constant for longer durations~\cite{mattu2025differential}. This allows efficient precoder design. In this paper, we design a precoder~\cite{masouros2012interference} using the QR-decomposition~\cite{horn2012matrix} of the channel matrix to reduce the doubly spread channel model to a Gaussian multiple access channel~\cite{harshan2011two}. 
This reduction makes it possible to
employ trellis coded modulation (TCM) to encode information symbols. In TCM, the minimum squared distance of the code compensates for the increased power of the larger constellation and the TCM encoder does not introduce additional symbols, thereby maintaining the interference levels of the uncoded system.

Numerical results demonstrate that the proposed TCM encoded MUB scheme with Zak-OTFS in the DD domain achieves good bit-error performance. The results also show that the proposed scheme achieves better performance 
than
orthogonal frequency domain multiplexing (OFDM) and OTFS 1.0 using the conventional faster-than-Nyquist scheme for the same spectral efficiency.

\textit{Notation:} $x$ denotes a complex scalar, $\mathbf{x}$ denotes a vector with $n$th entry $\mathbf{x}[n]$, and $\mathbf{X}$ denotes a matrix. $(\cdot)^{\mathsf{H}}$ denotes complex conjugate transpose and $(\cdot)^{\top}$ denotes transpose. Calligraphic font $\mathcal{A}$ denotes sets. $\mathbb{C}$ denotes the set of all complex numbers. 
$\mathsf{U}[x, y)$ denotes the continuous uniform random variable with limits $x$ (inclusive) and $y$ (exclusive). $\mathbf{I}_{M\times M}$ denotes the identity matrix of size $M \times M$ and $\mathbf{1}_{M\times N}$ denotes the all-ones matrix of size $M \times N$. $\mathbf{F}_M$ denotes the $M$-point discrete Fourier transform (DFT) matrix of size $M \times M$, $\otimes$ denotes Kronecker product, and $\mathsf{vec}(\cdot)$ denotes column-wise vectorization. $\Vert\cdot\Vert_2$ denotes the 2-norm of a vector and $\vert\cdot\vert$ denotes the absolute value.
\vspace{-0mm}

%% file: sys_model.tex
\section{System Model}
\label{sec:sys_model}

In Zak-OTFS, each information symbol is mounted on a \emph{pulsone} which is a pulse train modulated by a tone \cite{bitspaper1, bitspaper2}. A Zak-OTFS frame consists of $MN$ DD bins, where $M$ and $N$ are the number of delay and Doppler bins, respectively. The resolution along the delay axis is $\tau_p/M$ and that along Doppler axis is $\nu_p/N$, where $\tau_p = 1/\nu_p$ is called the delay period which is the inverse of the Doppler period. The frame bandwidth is $B=M\nu_p$ and the frame time is $T=N\tau_p$. $BT=MN$ information symbols drawn from a constellation alphabet $\mathcal{A}$, (e.g., 4-QAM) are mounted on these $MN$ DD bins by modulating each of the $MN$ pulsones. 
The Zak-OTFS input-output (I/O) relation is given by~\cite{bitspaper2}:
\begin{align}
    \mathbf{y} = \mathbf{H}\mathbf{x} + \mathbf{n},
    \label{eq:sys_model}
\end{align}
where $\mathbf{y} \in \mathbb{C}^{MN\times 1}$ is the vector of received symbols in the DD domain, $\mathbf{H}\in \mathbb{C}^{MN\times MN}$ is the end-to-end channel matrix, $\mathbf{x}\in\mathbb{C}^{MN\times 1}$ is the vector of transmitted symbols in the time-domain, and $\mathbf{n}\in \mathbb{C}^{MN\times 1}$ is the additive Gaussian noise. In the remainder of this paper, we assume the channel matrix $\mathbf{H}$ is perfectly known at the transmitter. In practice, such channel knowledge can be obtained by estimating the channel at the receiver from the response to pilot symbols~\cite{Aug2024paper,preamblepaper} and feeding back the estimated channel~\cite{masouros2012interference} to the transmitter.

\subsection{Preliminaries}
In this Subsection we present the system model for conventional faster-than-Nyquist schemes employed for improving spectral efficiency in OFDM~\cite{rusek_ftn, calavolpe_ftn_tf_2lattice_poweralloc_turbo, zhang_ftn_tf_hexalattice_isioptmzn} and MC-OTFS~\cite{hong2024precoded}.

\subsubsection{OFDM}
Let $\mathbf{X}_{\mathsf{TF}} \in \mathbb{C}^{M\times N}$ denote $MN$ information symbols in the time-frequency (TF) domain with $M$ frequency bins and $N$ time bins. The information symbols are converted to time domain via~\cite{Tse2005}:
\begin{align}
    \label{eq:ofdm_ftn1}
    \mathbf{s}_{t} = \mathsf{vec}(\mathbf{F}_{M}^{\mathsf{H}}\mathbf{X}_{\mathsf{TF}}) = (\mathbf{I}_{N\times N} \otimes \mathbf{F}_M^{\mathsf{H}})\mathsf{vec}(\mathbf{X}_{\mathsf{TF}}).
\end{align}
Let $\mathbf{H}_t(\beta)$ denote the time-domain channel matrix~\cite[Eq. (9)]{hong2024precoded}. The inter-symbol interference introduced from faster-than-Nyquist sampling is accounted for in $\mathbf{H}_t(\beta)$ via $\beta$. Then the TF channel matrix is represented as:
\begin{align}
    \label{eq:ofdm_ftn2}
    \mathbf{H}_{\mathsf{TF}} = (\mathbf{I}_{N\times N} \otimes \mathbf{F}_M)\mathbf{H}_t(\beta)(\mathbf{I}_{N\times N}\otimes \mathbf{F}_M^{\mathsf{H}}),
\end{align}
where the term $(\mathbf{I}_{N\times N} \otimes \mathbf{F}_M)$ follows from DFT operation at the receiver. The system model becomes:
\begin{align}
    \label{eq:ofdm_ftn3}
    \mathbf{y}_{\mathsf{TF}} = \mathbf{H}_{\mathsf{TF}}\mathbf{x}_{\mathsf{TF}} + \mathbf{n}_{\mathsf{TF}},
\end{align}
where $\mathbf{n}_{\mathsf{TF}}$ is additive noise, $\mathbf{x}_{\mathsf{TF}} = \mathsf{vec}(\mathbf{X}_{\mathsf{TF}})$, and $\mathbf{y}_{\mathsf{TF}} \in \mathbb{C}^{MN \times 1}$ denotes the received TF domain vector.

\subsubsection{OTFS 1.0}
Let $\mathbf{X}_{\mathsf{DD}} \in \mathbb{C}^{M\times N}$ denote $MN$ information symbols in the DD domain.
The information symbols are converted to time domain via inverse discrete Zak transform (IDZT)~\cite{hong2024precoded}:
\begin{align}
    \label{eq:otfs_ftn1}
    \mathbf{s}_{t} = \mathsf{vec}(\mathbf{X}_{\mathsf{DD}}\mathbf{F}_{N}^{\mathsf{H}}) = (\mathbf{F}_N^{\mathsf{H}} \otimes \mathbf{I}_{M\times M})\mathsf{vec}(\mathbf{X}_{\mathsf{DD}}).
\end{align}
The DD channel matrix is represented as:
\begin{align}
    \label{eq:otfs_ftn2}
    \mathbf{H}_{\mathsf{DD}} = (\mathbf{F}_N \otimes \mathbf{I}_{M\times M})\mathbf{H}_t(\beta)(\mathbf{F}_N^{\mathsf{H}} \otimes \mathbf{I}_{M\times M}),
\end{align}
where the term $(\mathbf{I}_{N\times N} \otimes \mathbf{F}_M)$ follows from DFT operation at the receiver. The system model becomes:
\begin{align}
    \label{eq:otfs_ftn3}
    \mathbf{y}_{\mathsf{DD}} = \mathbf{H}_{\mathsf{DD}}\mathbf{x}_{\mathsf{DD}} + \mathbf{n}_{\mathsf{DD}},
\end{align}
where $\mathbf{n}_{\mathsf{DD}}$ is additive noise, $\mathbf{x}_{\mathsf{DD}} = \mathsf{vec}(\mathbf{X}_{\mathsf{DD}})$, and $\mathbf{y}_{\mathsf{DD}} \in \mathbb{C}^{MN \times 1}$ denotes the received DD domain vector.

%% file: prop_method.tex
\section{Proposed Method}
\label{sec:prop_method}
The Zak-OTFS I/O relation in \eqref{eq:sys_model}
changes with the geometry of the reflectors, which is slow in practice%
~\cite{bitspaper1, bitspaper2}. This 
enables the following precoder design.

\subsection{Precoder Design}
\label{subsec:precoder_design}
The channel in Zak-OTFS introduces spreading. $P$ copies of each transmitted symbol are received for a $P$ path channel. To mitigate the effect of this spread, we design the precoder as described below.

The QR-factorization~\cite{horn2012matrix} of the hermitian of the channel matrix in~\eqref{eq:sys_model} is:
\begin{align}
    \label{eq:precoder1}
    \mathbf{H}^{\mathsf{H}} = \mathbf{Q}\mathbf{R},
\end{align}
where $\mathbf{Q}\in \mathbb{C}^{MN\times MN}$ is a unitary matrix and $\mathbf{R}$ is a lower (or upper) triangular matrix. The I/O relation from~\eqref{eq:sys_model} can be represented as:
\begin{align}
    \label{eq:precoder2}
    \mathbf{y} = \mathbf{R}^\mathsf{H}\mathbf{Q}^{\mathsf{H}}\mathbf{x} + \mathbf{n}.
\end{align}
We precode $\mathbf{x}$ with the matrix $\mathbf{Q}$, i.e, $\mathbf{x} = \mathbf{Q}\mathbf{x}'$, and~\eqref{eq:precoder2} simplifies to:
\begin{align}
    \label{eq:precoder3}
    \mathbf{y} = \mathbf{R}^{\mathsf{H}}\mathbf{x}' + \mathbf{n},
\end{align}
since $\mathbf{Q}^{\mathsf{H}}\mathbf{Q} = \mathbf{I}_{MN\times MN}$. At the receiver, a minimum mean square error (MMSE) matrix is constructed as:
\begin{align}
    \label{eq:precoder4}
    \mathbf{W} = (\mathbf{R}\mathbf{R}^{\mathsf{H}}+\sigma^2\mathbf{I})^{-1}\mathbf{R},
\end{align}
where $\sigma^2$ is the variance of the additive noise. Finally, the receiver computes:
\begin{align}
    \label{eq:precoder5}
    \mathbf{y}' = \mathbf{W}\mathbf{y}.
\end{align}
The precoding operation at the transmitter and combining operation at the receiver together reduce the input-output relation in~\eqref{eq:sys_model} to a Gaussian multiple access model. We describe how we communicate more than $MN$ symbols which makes use of this reduction.

\subsection{Mounting Info. Symbols on Mutually Unbiased Bases}
\label{subsec:mounting_symbols}

Let $\mathbf{x}_1', \mathbf{x_2}' \in \mathcal{A}^{MN \times 1}$ be two information vectors. Let $\mathbf{S}_1, \mathbf{S}_2 \in \mathbb{C}^{MN \times MN}$ denote two sets of bases. The bases each span the $MN$-dimensional space with the property that they are unbiased with respect to each other. Mathematically, the bases $\mathbf{S_1}, \mathbf{S_2}$ satisfy \cite{Aug2024paper}:
\begin{align}
    \label{eq:mut_unb_basis}
    \vert \mathbf{S}_i^{\mathsf{H}}\mathbf{S}_j\vert = \begin{cases}
        \mathbf{I}_{MN\times MN}, \ \ \ \ \ \quad \quad \text{if } i=j \\
        \frac{1}{\sqrt{MN}}\mathbf{1}_{MN \times MN}, \quad \text{if } i\neq j
    \end{cases}.
\end{align}

We mount $2MN$ information symbols on the Nyquist-grid\footnote{Here, by Nyquist grid, we mean the original OTFS grid, which has $MN$ degrees of freedom, with bandwidth $B=M\nu_p$ and time $T=N\tau_p$.}:
\begin{align}
    \label{eq:ftn_sys_model}
    \mathbf{y}' = \mathbf{W}\mathbf{H}\big(\sqrt{\alpha}\mathbf{Q}\mathbf{S}_1\mathbf{x}_1' + \sqrt{1-\alpha}\mathbf{Q}\mathbf{S}_2\mathbf{x}_2'\big) + \mathbf{W}\mathbf{n},
\end{align}
where 
the parameter $\alpha$ allocates power between the two frames%
\footnote{Energy in the first frame is $\alpha$ and in the second frame is $1-\alpha$, since $\mathbf{x}$ is drawn from a unit energy constellation. The total transmit energy is still unity.}. Notice that using the system model as defined above \textit{does not} incur energy, time, or bandwidth expansion.
Using the fact that $\mathbf{W}\mathbf{H}\mathbf{Q} \approx \mathbf{I}_{MN \times MN}$, the input-output relation can be approximated as:
\begin{align}
    \label{eq:ftn_pre_sys_model}
    \mathbf{y}' \approx \sqrt{\alpha}\mathbf{S}_1\mathbf{x}_1' + \sqrt{1-\alpha}\mathbf{S}_2\mathbf{x}_2' + \mathbf{n}'.
\end{align}
Note that we mount the information symbols on the mutually unbiased bases and information symbols used in $\mathbf{x}_1$ and $\mathbf{x}_2$ need not be from a uniquely decodable constellation set \cite{harshan2011two}.

\subsection{Detection of Info. Symbols}
\label{subsec:detection_symbols}
At the receiver, the vector $\mathbf{y} \in \mathbb{C}^{MN \times 1}$ is received. The received $\mathbf{y}$ is combined to get $\mathbf{y}'$ (see~\eqref{eq:precoder5}). To detect $\mathbf{x}_1'$ and $\mathbf{x}_2'$ from $\mathbf{y}'$, we proceed as follows.
For detecting the $i$th frame ($i \in \{1, 2\}$), we perform pre-multiplication by the complex conjugate transpose of the basis matrix $\mathbf{S}_{i}$:
\begin{align}
    \label{eq:first_frame_eq}
    \mathbf{S}_{i}^{\mathsf{H}}\mathbf{y}' = \beta_{i}\mathbf{x}_{i}' + \tilde{\mathbf{n}}_{i},~i \in \{1,2\},
\end{align}
where $\tilde{\mathbf{n}}_{i} = \mathbf{S}_{i}^{\mathsf{H}}(\beta_{j}\mathbf{S}_{j}\mathbf{x}_{j}' + \mathbf{n}'), j \in \{2, 1\}$ is the resulting noise for $\beta_{1} = \sqrt{\alpha},~\beta_{2} = \sqrt{1-\alpha}$. Since matrices $\mathbf{S}_i, i=1, 2$ are chosen to be orthonormal, there is no noise enhancement. To recover information symbols $\mathbf{x}_i' \in \mathbb{C}^{MN \times 1}$, we perform and element-wise detection in Gaussian channel as:
\begin{align}
    \label{eq:first_frame_det}
    \hat{\mathbf{x}}'_i[q] = \underset{s \in \mathcal{A}}{\arg\min} \Vert \mathbf{S}_i^{\mathsf{H}}\mathbf{y}'[q] - \beta_is\Vert^2,
\end{align}
where $q = 0, 1, \cdots, MN-1$. Next, for detection of second frame, we cancel the contribution of the detected frame from the received frame as:
\begin{align}
    \label{eq:first_frame_rem}
    \bar{\mathbf{y}} = \mathbf{y}' - \beta_i\mathbf{S}_i\hat{\mathbf{x}}_i'.
\end{align}
To detect the second frame, the same Gaussian detection is performed as:
\begin{align}
    \label{eq:second_frame_det}
    \hat{\mathbf{x}}'_j[q] = \underset{s \in \mathcal{A}}{\arg\min} \Vert \mathbf{S}_j^{\mathsf{H}}\bar{\mathbf{y}}[q] - \beta_js\Vert^2,
\end{align}
where $q = 0, 1, \cdots, MN-1$.

\subsection{Trellis Coded Modulation}
\label{subsec:tcm}
From \eqref{eq:first_frame_eq} it is clear that the noise floor increases, since the effective noise is the sum of the additive Gaussian noise and the term $\mathbf{S}_i^{\mathsf{H}}\beta_j\mathbf{S}_j\mathbf{x}_j'$. This decreases the effective signal to interference plus noise ratio. The detector described in \eqref{eq:first_frame_det} or \eqref{eq:second_frame_det} therefore would have poor bit-error performance. To lift the information symbols well above noise level we use coding, specifically TCM, which ensures that the coded symbols do not create additional interference to the other frame, i.e., the number of uncoded symbols and coded symbols is the same albeit the coded symbols come from a higher constellation alphabet.
The detection of information symbols follows that described in Sec. \ref{subsec:detection_symbols} with the only change being that instead of using detection in Gaussian channel as shown in \eqref{eq:first_frame_det}, we use a Viterbi decoder. 

\subsection{Turbo Iterations}
\label{subsec:turbo}
The first frame is decoded in the presence of the second frame. Subsequently, the decoded frame is re-encoded and subtracted from the received frame before the second frame is decoded. The performance can be improved if more rounds of interference cancellation and decoding are performed. This is termed as turbo iterations. Specifically, the decoded second frame is re-encoded and removed from the received frame and the first frame is decoded again. The decoded first frame is re-encoded and subtracted from the received frame before decoding the second frame. This forms one turbo iteration. Many turbo iterations could be performed until the performance improvement between two successive turbo iterations becomes insignificant.

\subsection{Maximizing Effective Rate}
\label{subsec:eff_rate}
\begin{figure}
    \centering
    \includegraphics[width=0.8\linewidth]{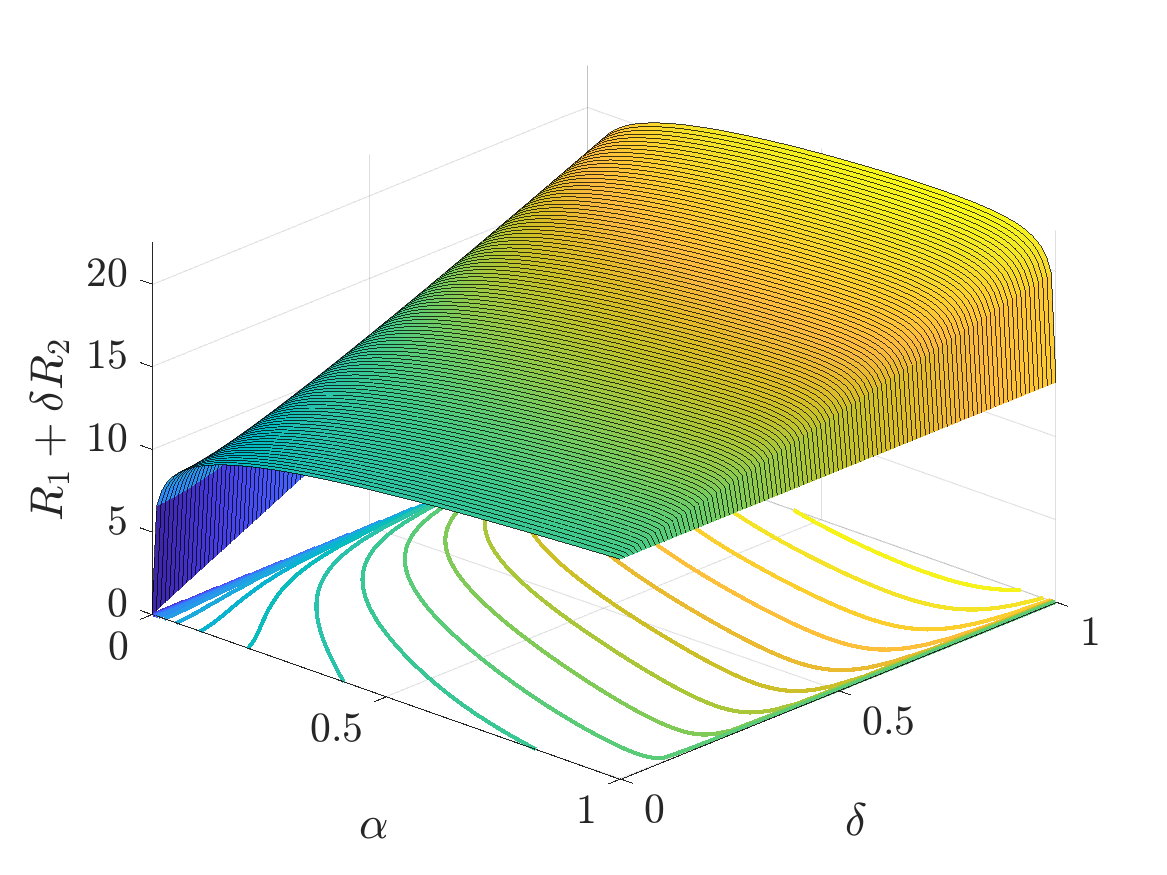}
    \caption{Effective rate as a function of power distribution $\alpha$ and second frame sparsity $\delta$ for SNR of $40$ dB. $M=31, N=37, d_{\mathsf{free}}=\sqrt{20}$. For a given $\delta$, as $\alpha$ increases the effective rate increases, but dips when $\alpha=1$, which corresponds to single frame transmission. For a given $\alpha$, the effective rate increases with $\delta$.}
    \label{fig:rate_vs_alpha_vs_delta}
    \vspace{-2mm}
\end{figure}

In this Subsection, we derive the effective rate as a function of the power distribution between the frames and sparsity of the second frame. Let $P_1 = \alpha P$ denote the energy in the first frame and $P_2 = (1-\alpha)P$ denote the energy in the second frame. Let $\nicefrac{P}{\sigma^2}$ denote the effective signal-to-noise ratio. The signal to interference plus noise ratio for the first frame is:
\begin{align}
    \mathsf{SINR}_1 = \frac{P_1}{\sigma^2 + \frac{\delta P_2}{MN}},
\end{align}
where $\delta \in [0, 1]$ denotes the sparsity of the second frame. The denominator follows from the fact that the second frame acts as interference to the first frame and that the interference is flat (see~\eqref{eq:mut_unb_basis}). Before we detect the second frame, we detect the first frame and subtract it from the received frame. Let $P_{s_1}$ denote the symbol error probability for the first frame. Then the signal to interference plus noise ratio for the second frame is:
\begin{align}
    \mathsf{SINR}_2 = \frac{P_2}{\sigma^2 + \frac{P_1P_{s_1}}{MN}}.
\end{align}
Given the $\mathsf{SINR}_1$, $P_{s_1}$ is given by~\cite{ungerboeck1982channel}:
\begin{align}
    P_{s_1} = Q\Big(\sqrt{2d_{\mathsf{free}}\mathsf{SINR}_1}\Big),
\end{align}
where $Q(\cdot)$ denotes the $Q$-function, $d_{\mathsf{free}}$ is the free Euclidean distance of the TCM trellis. From the Shannon formula, the rate can be written as:
\begin{align} 
    R_1 &= \log_2(1+\mathsf{SINR}_1) \\
    R_2 &= \log_2(1+\mathsf{SINR}_2).
\end{align}
The goal is to now choose $P_1$ and $P_2$, or equivalently $\alpha$, to maximize the effective rate $R_1+\delta R_2$. Fig.~\ref{fig:rate_vs_alpha_vs_delta} shows the plot of effective rate as a function of $\alpha$ and $\delta$. The optimal choice for $\alpha$ is a value close to, but not equal to, $1$.

\subsection{Spectral Efficiency \& Performance Tradeoff}
\label{subsec:choosing_delta}
The input-output relation is given by~\eqref{eq:ftn_pre_sys_model}. Suppose the detection for frame $\mathbf{x}_1'$ is carried out first. Then, the signal energy is:
\begin{align}
    \label{eq:choose_delta1}
    E_s = \Vert \sqrt{\alpha}\mathbf{S}_x\mathbf{x}_1'\Vert_2^2 = \vert\alpha\vert = \alpha,
\end{align}
since the information symbols are chosen from a unit energy constellation and the basis is orthonormal. The interference plus noise power is:
\begin{align}
    \label{eq:choose_delta2}
    E_{n} = \Vert\sqrt{1-\alpha}\mathbf{S}_2\mathbf{x}_2'+\mathbf{n}'\Vert_2^2 = (1-\alpha)\delta + \sigma^2,
\end{align}
where the last equality follows from the fact that $\mathbf{x}_2'$ is $\delta$-sparse and $\sigma^2$ is the variance of noise. For detection of $\mathbf{x}_1'$, we would require $E_s$ to be greater than $E_n$. Let $\gamma>1$ be a real number, then:
\begin{align}
    \label{eq:choose_delta3}
    E_s \geq \gamma E_n \implies \alpha \geq \gamma((1-\alpha)\delta + \sigma^2) \implies  \delta \leq \frac{\alpha-\gamma\sigma^2}{\gamma(1-\alpha)}.
\end{align}
Hence, $\delta$ cannot be arbitrarily large. The inverse relation between $\delta$ and $\gamma$ implies that increasing $\delta$ leads to poor performance for given $\sigma^2$ and $\alpha$. This puts a limit on the maximum achievable spectral efficiency much like the Mazo limit for the conventional faster-than-Nyquist signaling~\cite{mazo1975faster}.

%% file: results.tex
\begin{table}[!t]
    \centering
    \caption{Power-delay profile of Veh-A channel model}
    \begin{tabular}{|c|c|c|c|c|c|c|}
         \hline
         Path index $i$ & 1 & 2 & 3 & 4 & 5 & 6 \\
         \hline
         Delay $\tau_i (\mu s)$ & 0 & 0.31 & 0.71 & 1.09 & 1.73 & 2.51 \\
         \hline
         Relative power (dB) & 0 & -1 & -9 & -10 & -15 & -20 \\
         \hline
    \end{tabular}
    \label{tab:veh_a}
\end{table}

\begin{figure*}
    \centering
    \begin{subfigure}{0.49\linewidth}
        \includegraphics[width=0.9\textwidth]{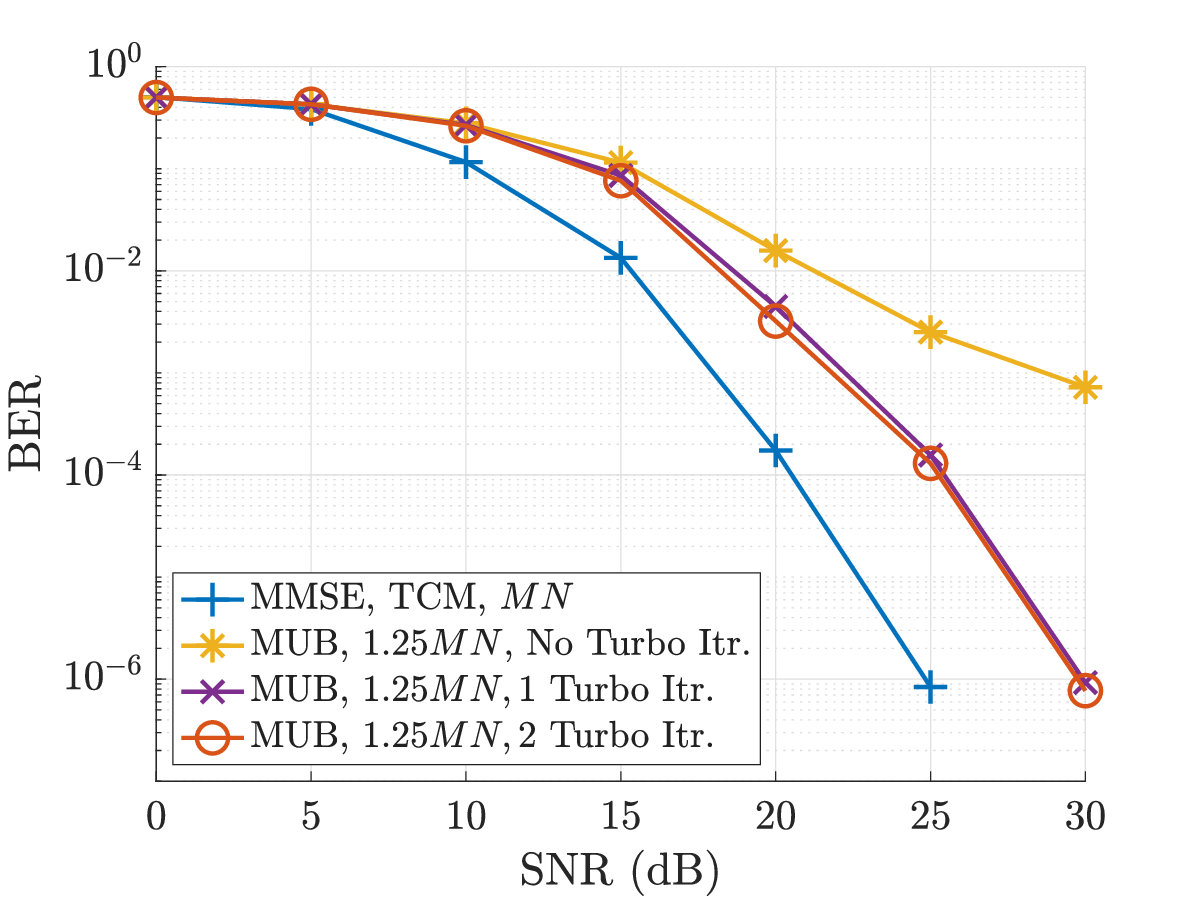}
        \caption{$1.25MN$.}
        \label{fig:1_p_25_turbo}
    \end{subfigure}
    \begin{subfigure}{0.49\linewidth}
        \includegraphics[width=0.9\textwidth]{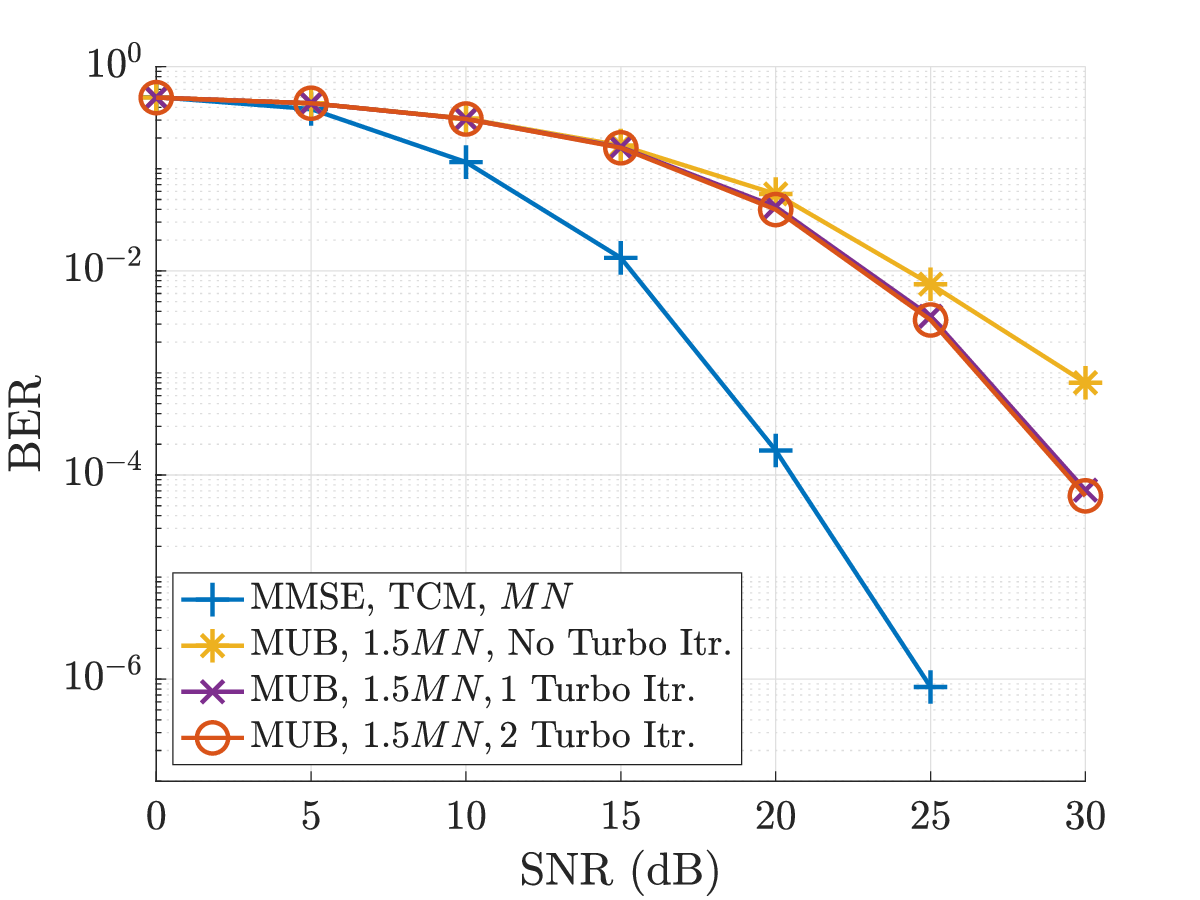}
        \caption{$1.5MN$}
        \label{fig:1_p_5_turbo}
    \end{subfigure}
    \caption{Bit-error performance of the proposed MUB scheme as a function of SNR for $1.25MN$ and $1.5MN$ symbols. Sinc pulse shaping, a $6$-path Vehicular-A channel per Table~\ref{tab:veh_a}, and parameters $M = 31$, $N = 37$, $\nu_p = 30$ kHz. Turbo iterations improve performance. No significant improvement beyond $2$ turbo iterations.}
    \vspace{-5mm}
    \label{fig:ber_turbo}
\end{figure*}

\begin{figure*}[t]
    \centering
    \begin{subfigure}{0.49\linewidth}
        \includegraphics[width=0.9\textwidth]{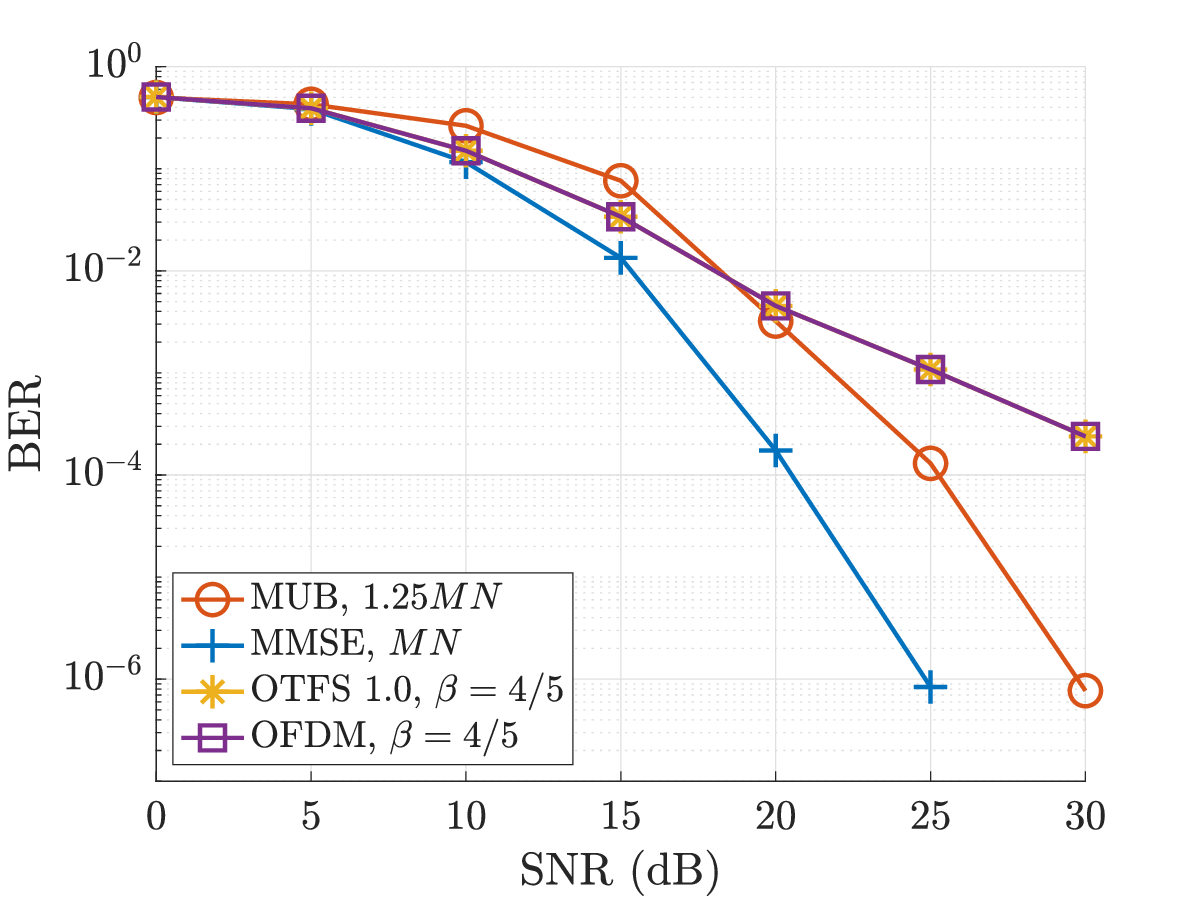}
        \caption{$1.25MN$.}
        \label{fig:1_p_25_MN_comp}
    \end{subfigure}
    \begin{subfigure}{0.49\linewidth}
        \includegraphics[width=0.9\textwidth]{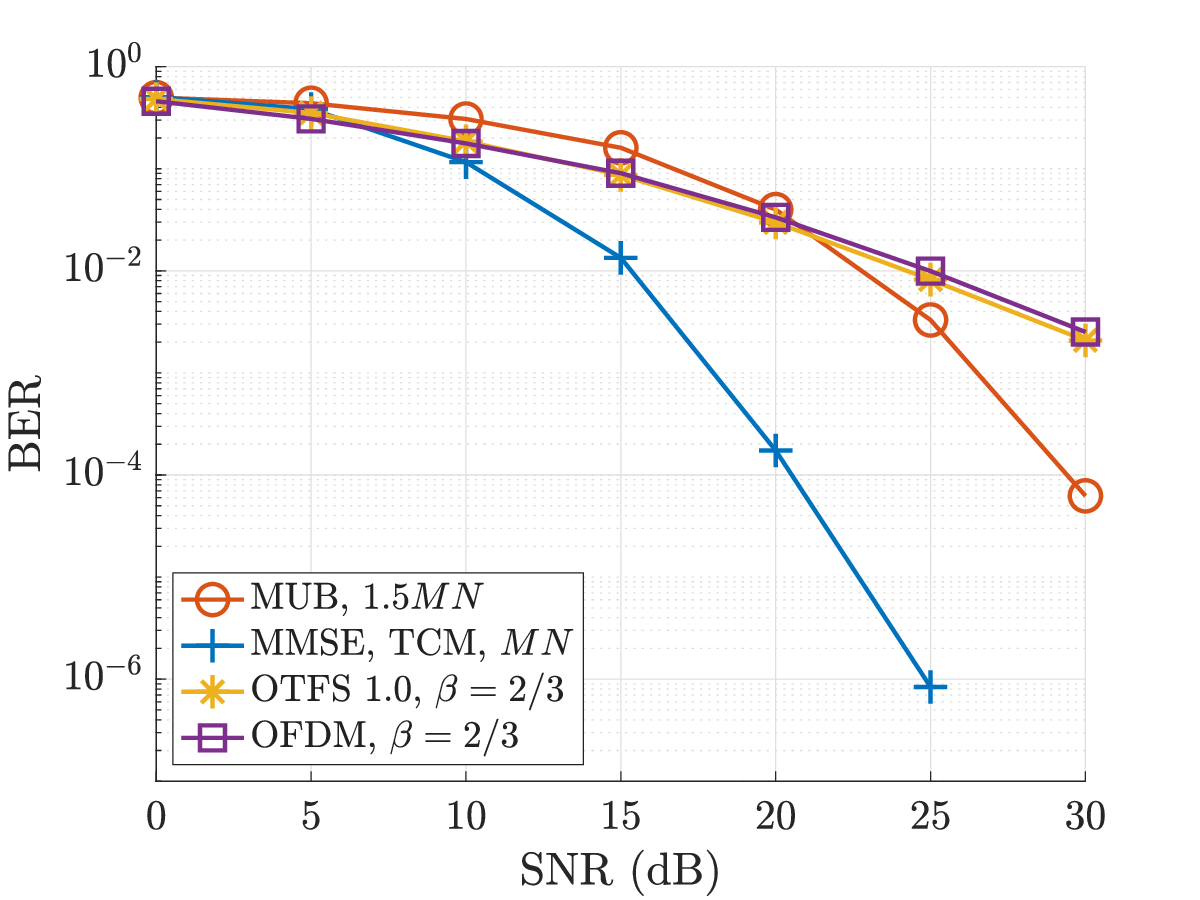}
        \caption{$1.5MN$}
        \label{fig:1_p_5_MN_comp}
    \end{subfigure}
    \caption{Bit-error performance of the proposed MUB scheme as a function of SNR for $1.25MN$ and $1.5MN$ symbols. Sinc pulse shaping, a $6$-path Vehicular-A channel per Table~\ref{tab:veh_a}, and parameters $M = 31$, $N = 37$, $\nu_p = 30$ kHz. The performance of OFDM and OTFS 1.0 begin to floor at high SNRs, while the proposed MUB approach based on Zak-OTFS shows better BER performance.}
    \vspace{-5mm}
    \label{fig:ber_comp}
\end{figure*}

\section{Numerical Results}
\label{sec:results}
In this Section, we present the numerical simulations to evaluate the performance of the proposed faster-than-Nyquist signaling scheme. For all the simulations we consider Zak-OTFS modulation with $M=31, N=37$ and the Doppler period $\nu_p = 1/\tau_p = 30$ kHz. For the channel, we consider a Vehicular-A channel \cite{veh_a} with power delay profile given by Table \ref{tab:veh_a}. Sinc pulse shaping is used at the transmitter and receiver. Doppler of each path is generated using $\nu_i = \nu_{\max}\cos(\theta_i)$ with $\theta_i \sim \mathsf{U}[-\pi, \pi)$ and $\nu_{\max} = 815$ Hz. For the MUB scheme, we consider the first frame ($\mathbf{x}_1$) to be a full frame containing $MN$ symbols and the superimposed second frame ($\mathbf{x}_2$) to be sparse (either $0.25MN$ or $0.5MN$ symbols) and detect the full frame first. We TCM encode both the frames. For TCM encoding, we use the following rule:
\begin{align}
    \label{eq:4_qam_16_qam_encoding}
    s = (-1)^{(\mathbf{g}_{11}^\top\mathbf{b})_2} - 3(-1)^{(\mathbf{g}_{12}^\top\mathbf{b})_2}
\end{align}
where $\mathbf{g}_{11} = [0 \ 1 \ 0]^\top, \mathbf{g}_{12} = [1 \ 1 \ 1]^\top$ are the generator polynomials, $\mathbf{b} \in \{0, 1\}^{n \times 1}$ is a binary vector to the input of the encoder, and $n = 3$ is the sliding window length. The output of TCM encoder takes values in the set $\{-2, -1, 1, 2\}$. Complex information symbols are constructed by treating TCM encoded output at even indices as real part and at the odd indices as corresponding imaginary part. The TCM encoder converts 4-QAM symbols (i.e., 2 bits) to amplitude modulated 16-QAM symbols (i.e., 4 bits)\footnote{In this paper, as a proof of concept, we use this design for the TCM encoder. However, optimizing the code design is an important direction for future research.}. For Zak-OTFS, OTFS 1.0, and OFDM, we use the same precoding technique described in Sec.~\ref{subsec:precoder_design}. Further, for fair comprision we use the same TCM encoding for all modulation schemes. The power distribution between the frames is fixed by choosing $\alpha=0.9$.


\subsection{Effect of Turbo Iterations}
\label{subsec:turbo_effect}
Fig.~\ref{fig:ber_turbo} shows the effect of turbo iterations on the bit-error performance of the proposed MUB receiver. Fig~\ref{fig:ber_turbo}(\subref{fig:1_p_25_turbo}) shows the bit-error rate (BER) when $\delta = 0.25$. The performance with regular minimum mean square error (MMSE) with TCM encoded $MN$ symbols is also added for comparison. It is seen that without turbo iteration, the performance begins to floor at high SNR values. However, with $1$ turbo iteration, it seen that the performance is much better at high SNR values. Adding a second turbo iteration doesn't improve the performance significantly. Similar observations hold when $\delta = 0.5$ in Fig.~\ref{fig:ber_turbo}(\subref{fig:1_p_5_turbo}). A single turbo iteration is sufficient to get an order of BER improvement at high SNR values with slight improvement with $2$ turbo iterations. In the following simulations, we present results with $2$ turbo iterations.

\subsection{Comparison With Other Schemes}
\label{subsec:16_qam}
Fig.~\ref{fig:ber_comp} compares the performance of the proposed MUB scheme against those of OFDM (see~\eqref{eq:ofdm_ftn3}) and OTFS 1.0 (see~\eqref{eq:otfs_ftn3}) for $\delta = 0.25, 0.5$. Note that, for OFDM and OTFS 1.0 we use the conventional faster-than-Nyquist signaling (see Fig.~\ref{fig:ftn_comp}(\subref{fig:conv_approach})). For the MUB scheme, when $\delta = 0.25$ we transmit $1.25MN$ symbols in bandwidth $B$ and time $T$. This implies a compression ratio of $\nicefrac{1}{1.25} = \nicefrac{4}{5}$. Therefore for comparison, in OFDM and OTFS 1.0 we choose $\beta=\nicefrac{4}{5}$ (see Fig.~\ref{fig:ftn_comp}(\subref{fig:conv_approach})). Similarly, when $\delta=0.5$, we choose $\beta=\nicefrac{2}{3}$. In Fig.~\ref{fig:ber_comp}(\subref{fig:1_p_25_MN_comp}) when $\delta=0.25$, the performance of the proposed MUB scheme is worse than the MMSE performance with $MN$ symbols by only $5$ dB. On the other hand, the performance of OFDM and OTFS 1.0 is better at low SNRs but floors at high SNRs. Similar trends are observed when $\delta=0.5$ in Fig.~\ref{fig:ber_comp}(\subref{fig:1_p_5_MN_comp}). The performance of the proposed MUB is worse by a factor of about $8$ dB for a BER of $10^{-2}$ while the OFDM and OTFS 1.0 are worse by about $10$ dB. The gap between the performance of MMSE and OFDM \& OTFS 1.0 only grows as SNR increases.

%% file: conclusion.tex
\section{Conclusion}
\label{sec:conclusion}
In this paper, we proposed MUB scheme in Zak-OTFS to improve spectral efficiency in channels with mobility and delay spread. The proposed method did not require changing the sampling frequency resulting in efficient implementation with existing hardware. The proposed scheme leveraged the slow variation of the DD channel in Zak-OTFS to construct an efficient precoder that mitigated the effect of the doubly-spread channel. Employing TCM encoding and using turbo iterations at the receiver enabled the proposed MUB scheme to achieve superior performance when compared to OFDM and OTFS 1.0 for the same spectral efficiency improvement. 